\documentclass[aps,pre,superscriptaddress,twocolumn,amsmath,amssymb]{revtex4}
\usepackage{graphicx}
\usepackage{dcolumn}
\usepackage{bm}

\begin{document}

\title{Econophysics studies in Estonia}

\author{M. Patriarca}
 \affiliation{National Institute of Chemical Physics and Biophysics, 
  R\"avala 10, Tallinn 15042, Estonia}
  \affiliation{IFISC, Instituto de F\'isica Interdisciplinar y Sistemas Complejos (CSIC-UIB), 
  E-07122 Palma de Mallorca, Spain}

\author{E. Heinsalu}
 \affiliation{National Institute of Chemical Physics and Biophysics, 
  R\"avala 10, Tallinn 15042, Estonia}
  \affiliation{IFISC, Instituto de F\'isica Interdisciplinar y Sistemas Complejos (CSIC-UIB), 
  E-07122 Palma de Mallorca, Spain}

\author{R. Kitt}
\affiliation{Institute of Cybernetics, Tallinn University of Technology, Akadeemia tee 21, 12618, Tallinn, Estonia}
\affiliation{Swedbank AS, Liivalaia 12, 15038, Tallinn, Estonia}

\author{J. Kalda}
\affiliation{Institute of Cybernetics, Tallinn University of Technology, Akadeemia tee 21, 12618, Tallinn, Estonia}


%

%


\maketitle


\section{Introduction}


The term ``econophysics'', proposed in 1995 in Kolkata by E.~H.~Stanley, celebrates 15 years; 
however, the studies of economic systems using the tools of statistical physics initiated much earlier.
Indeed, V.~Pareto formulated his famous law of income distribution already in 1897 \cite{Pareto}, and only three years later, L.~Bachelier put forward the random walk model as the fundamental model for financial time-series \cite{Bachelier1990b}.
Further, E.~Majorana pointed out that statistical mechanics is a tool that can be applied also in social sciences \cite{Majorana1942a}.
In 1963 B.~Mandelbrot found that the time-series of cotton price undergoes large fluctuations \cite{Mandelbrot1963a}.
The studies of econophysics have developed at an ever accelerating rate since 80's, particularly fast after the adoption of the very term,  
until drawing the attention of Estonian physicists at the beginning of the new century.
The present paper is aimed to provide a short overview of the econophysical research in Estonia, which thus far has
resulted in more than 15 research papers.

Economy is a very good example of complex systems: numerous building blocks interact with each other and form a system with qualitatively new properties, intermittent and scale-invariant behaviour. 
However, in the case of economy, already the building blocks --- humans performing economical activities ---  are not simple entities.
This is unlike to what is observed in the case of ``simple'' complex systems, such as sandpiles or turbulent flow. 
The physics of complex systems has taught us that even these ``simple'' complex systems can lead to a strong intermittency and are very difficult to study theoretically. 
So, it becomes clear that in econophysics, one can find examples of an ultimate complexity. 

What can be done in the case of such an extreme complexity? 
There is probably not much sense to make very detailed models, just because it is impossible to account for all the ``unreasonable players''. 
We are basically left with two options: first, we can make very robust models, which, although inaccurate, capture certain features of the economical dynamics. 
For example, one can introduce an ensemble of stochastically interacting traders to derive a power law scaling of market prices \cite{Lux1999}. 
Similarly, one can use the Pareto law for market participants and postulate optimal trading behaviour to derive the power law for the probabilities of large market movements \cite{Gabaix2003}. 
Alternatively, one can follow a (semi)empirical approach to devise general statistical descriptions of the spatio-temporal behaviour of markets. 
For instance, one can use Trade  and Quote databases to show the existence of two phases in the behaviour of financial markets: besides the equilibrium phase, there is also a phase where the most probable behaviour is either selling or buying \cite{Plerou2003}. Another example is provided by the study, which shows that upon properly accounting for the market capitalization of different economic sectors, the price-impact data can be collapsed into a single power law as a function of the transaction size \cite{Stanley2003}. 

In Estonia, both, the branch of robust models as well as that of the semi-empirical approach, are represented.
The corresponding studies are carried out, respectively, at the National Institute of Chemical Physics and Biophysics and at the Institute of Cybernetics at Tallinn University of Technology, both in Tallinn.


\section{Many-agent wealth exchange models}


The researchers at the National Institute of Chemical Physics and Biophysics have been
investigating {\it kinetic wealth-exchange models} (KWEM) of closed economy systems.
This type of models were independently introduced in different fields such as social sciences \cite{Angle1983a,Angle1986a,Angle2006a}, 
economics \cite{Bennati1988a,Bennati1988b,Bennati1993a}, and (econo)physics \cite{Chakraborti2000a,Slanina2004a}.
The International Meeting \emph{Econophys-Kolkata 1} \cite{Chatterjee2005b} hold in 2005
had an important role in integrating the knowledge accumulated that
far about the various models, bringing out the earlier works of John Angle \cite{Lux2005a-brief}
and Eleonora Bennati \cite{Patriarca2005a-brief}, and in unveiling the basic mechanism in action 
leading to the appearance of a power law tail in the stationary wealth distribution in KWEMs.

Let us  discuss the general structure and features of KWEMs.
In a KWEM the system is made up of $N$ agents,
whose status at a certain time $t$ is characterized by the
wealths $x_i(t) \ge 0$ ($i = 1,2,\dots, N$).
In simple models agents interact with each other pairwise: at every time step two randomly chosen agents $j$ and $k$ 
exchange an amount $\Delta x$ of wealth so that the total wealth is conserved.
The new values $x_j'$ and $x_k'$ after the exchange are ($x_j', x_k' \ge 0$)
\begin{eqnarray} \label{evol}
  x_j' &=& x_j - \Delta x \, ,
  \nonumber \\
  x_k' &=& x_k + \Delta x \, .
  \label{basic0}
\end{eqnarray}
The form of $\Delta x $ defines the underlying dynamics of the model;
in the most simple case it is a constant \cite{Bennati1988a,Bennati1988b,Bennati1993a}, or it can be a
function of $x_j$, $x_k$ and some parameter characterizing the agents \cite{Patriarca2010b}.
In more complex multi-agent interaction models the number of agents that enter each trade is $M > 2$.
Then the evolution law has a more general form $x_i' = x_i + \Delta x_i$, with $i=1,\dots,M$, $\sum_{i=1}^M \Delta x_i = 0$, 
and the $\Delta x_i$ depending somehow on the wealths $x_i$ of the $M$ interacting agents.

Typically in KWEMs there is solely one parameter characterizing the agents. 
For example, it can be an exchange parameter $\omega \in (0, 1]$
which defines the fraction of the wealth $x$ that enters the exchange process.
Equivalently with $\omega $ one can introduce the saving parameter, $\lambda = 1 - \omega$, representing the fraction of $x$ preserved during the exchange.
If the value of $\omega $ (or $\lambda$) is the same for all the agents, the model is referred to as {\it homogeneous}.
If the agents assume different values $\omega _i$ (or $\lambda _i$) then the model is called {\it heterogeneous}.
As studied in Ref.~\cite{Patriarca2007a}, the parameter $\omega$ (or $\lambda$) also determines the time scale of the relaxation process.
In Ref.~\cite{Patriarca2010b} a unified formulation of various types of KWEMs with pairwise interacting agents was provided.

For $\omega < 1$ (or $\lambda > 0$), the homogeneous KWEMs have the self-organizing property to converge toward a stable state with a wealth distribution 
which has a non-zero median, differently from a purely exponential distribution.
It is well fitted by a $\Gamma$-distribution, which describes real wealth distributions in the range of small and intermediate values 
of the wealth, representing most of the agents \cite{Yakovenko2009a}.
In Ref.~\cite{Patriarca2004b} it was shown that the
equilibrium wealth distribution $f_n(x)$ of the homogeneous KWEM proposed in Ref.~\cite{Chakraborti2000a}, defined by
$\Delta x = \omega ( \bar{\epsilon}x_j - \epsilon x_k) = (1 - \lambda) ( \bar{\epsilon}x_j - \epsilon x_k)$,
where $\epsilon$ and $\bar{\epsilon} = 1 - \epsilon$ are two uniform random numbers in $(0,1)$,
is very similar to the following $\mathrm{\Gamma}$-distribution,
\begin{eqnarray}
  \frac{\langle x \rangle}{n} \, f_n(x) 
   = 
   \left(\! 
      \frac {n \, x} {\langle x \rangle}
   \!\right)^{\!n-1}
   \frac {\mathrm{e}^{- n x/\langle x \rangle}} {\Gamma(n)}
   =
   \frac {\xi^{n-1} \mathrm{e}^{-\xi}} {\Gamma(n)} 
   \equiv 
   \gamma_n(\xi)   
   \, .
  \label{NGibbs}
\end{eqnarray}
Here $\xi = n \, x / \langle x \rangle$,
with $n(\lambda) =  (1 - 2 \lambda)/(1 - \lambda)$,
$\langle x \rangle = \sum_i x_i /N$ being the (constant) average wealth of the
system, and $\gamma_n(\xi)$ being the $\mathrm{\Gamma}$-distribution.
In Ref.~\cite{Patriarca2010b} a comparison between the wealth equilibrium distributions of various other KWEMs was presented.

Let us discuss the link between KWEMs and statistical physics.
The form (\ref{evol}) of the exchange law suggests an analogy with the energy transfer between molecules of a 
fluid \cite{Chatterjee2005b,Hayes2002a,Chatterjee2007b}.
Furthermore, the distribution (\ref{NGibbs}) is well known in statistical mechanics,
representing for example the canonical distribution of the molecular kinetic energy
of a gas in $D = 2 n$ dimensions if, following the equipartition theorem, 
the average kinetic energy is assumed to be
$\langle x \rangle = D T /2 = nT$, where $T$ is the temperature.
This direct link between KWEMs and statistical physics can be confirmed through a standard kinetic theory
approach \emph{a la Clausius}, which shows that 
homogeneous KWEM agents with saving parameter $\lambda$
behave dynamically as the molecules of a gas in $D(\lambda) =
2 \, n(\lambda)$ dimensions \cite{Chakraborti2008a}.

In Ref.~\cite{Patriarca2004a} it was shown that when all the agents are trying to save
as much as possible ($\lambda \to 1$),
the distribution (\ref{NGibbs}) tends to an egalitarian distribution,
i.e., in the end all agents have the same amount of wealth $\langle x \rangle$.

While the  $\mathrm{\Gamma}$-distribution provides a good
fitting to the stationary wealth distributions, 
it does not seem to represent the exact solution \cite{Repetowicz2005a}.
The search of the actual shape of the equilibrium
wealth distribution is still an active topic of
research \cite{Repetowicz2005a,Angle2006a,Chakrabarti2009a,Patriarca2010b}.

As mentioned, the real wealth distributions in the range of small and intermediate values 
of the wealth, representing most of the agents, are well fitted by a $\Gamma$-distribution \cite{Yakovenko2009a}.
Instead, the remaining agents with large values of wealth, forming a small part of the system
yet owning a significant fraction of the total wealth, are described by the Pareto power law. 
Such power law tails are reproduced by heterogeneous models.
Indeed, as discussed in Ref.~\cite{Patriarca2006c}, heterogeneous models with diversified parameters $\omega _i$ (or $\lambda _i$) provide both, the exponential or $\Gamma$-distribution shapes of real wealth distributions at small and intermediate values as well as the Pareto power law observed at larger values of wealth.

Thus, heterogeneity plays a crucial role in the generation of the power law tail:
a power law is produced by the overlap of the single agent wealth distributions, which are fitted by
$\Gamma$-distributions with different parameters.
In other words, the global wealth distribution can be resolved as a
mixture of partial wealth probability densities with exponential tails.
This has been suggested in Ref.~\cite{Chakraborti2009a} to represent only one example
of a more general mechanism leading to the appearance of a power law tail in many other complex systems,
including the Zipf law in linguistics, i.e., the power law observed
in the rank distribution of words in a written text.

In Ref.~\cite{Patriarca2007a} the wealth scale and the time scale of a KWEM have been investigated.
In a heterogeneous KWEM, both the wealth scale and the time scale of each agent $i$ is determined by the parameter $\lambda_i$ (or $\omega_i$).
An agent with saving parameter $\lambda_i$
has a relaxation time $\tau_i \propto 1/(1-\lambda_i)$. 
Therefore the relaxation time $\tau$ of the global heterogeneous system
is determined by the largest saving parameter $\Lambda = \mathrm{max}\{\lambda_i\}$ so that $\tau \propto 1/(1-\Lambda)$.
In a heterogeneous KWEM a power law can appear.
In this case a natural wealth scale is present in the stationary wealth distribution,
represented by the wealth cutoff $X$ at which the power law ends and
the wealth distribution goes to zero.
$X = \mathrm{max}\{ x_i \}$ represents the wealth of the richest agent.
The wealth cutoff $X$ and the saving parameter cutoff $\Lambda$ are
closely related to each other: the closer $\Lambda$ is to 1, the larger is $X$.
This goes well also with the proverb that rich is not the one who earns a lot, but the one who spends a little.

To summarize, KWEMs are minimal models of closed economies, 
in which the total amount of wealth $\sum_i x_i$ is constant and therefore can be used to describe systems where the
flow of money is  conserved.
However, it is noteworthy that even so KWEMs provide realistic shapes 
of the stationary wealth distributions, suggesting that there are two main factors underlying the wealth dynamics: the microscopic exchange mechanism between pairs of agents, and the heterogeneity of the agents.


\section{Semi-empirical studies of financial time-series}


Econophysics research at the Institute of Cybernetics  has been following predominantly the
semi-empirical approach. 
The first studies were triggered by the practical problem of optimal portfolio construction, which was faced by Robert Kitt at the Hansabank (formerly the largest bank in Estonia, now a part of Swedbank). The conventional econometric models to determine the optimal portfolio structure for many assets were found to yield unstable results in stock markets (cf. Ref.~\cite{kitt2004}). 
The first studies were relatively simple: they were aimed to test the ``regularity'' of the local markets by studying the self-affine scaling behaviour of the Baltic market indices and comparing these with the global ones (cf. Ref.~\cite{kitt2003}). 
In full agreement with the earlier studies, c.f. Ref.~\cite{Vandewalle1997}, it was found that the Baltic indices are statistically self-affine. However, the Hurst exponent values tended to be somewhat higher (0.6--0.7) than in the case of larger and longer established markets (ca 0.5). 
Therefore, it was hypothesized that new and far-from-equilibrium economies are characterized by stronger persistency in market movements.

Nowadays it is well understood that the self-affine behaviour is only a very specific case of scale-invariance. 
As argued above, the market dynamics is inherently extremely complex. 
Hence, in order to understand its statistical features, a statistical description, which is as generic as possible, is needed. 
Multifractality represents a significantly more generic class of scale-invariance than the Gaussian self-affine time-series and has been shown to be present in financial time-series \cite{Fisher1997,Vandewalle1998}. 
However, multifractality implies also the validity of a specific assumption, namely, the presence of a random multiplicative cascade. 
This is in a natural way satisfied in the case of turbulence: the frozen-in and conserved quantities (enstrophy, energy) are passed down to the next generation of vortices when a larger vortex splits into smaller ones. 
However, in the case of market fluctuations, there is no ``enstrophy", and no multiplicative cascades in the asset markets.
If the total amount of money is conserved this could lead to some kind of multifractality, if the market were to split into smaller fragments. 
However, there is  no such kind of market splitting in reality. 
Hence, it would be desirable to devise a statistical description of the scale-invariant time-series, which would be more generic than the multifractal formalism. Precisely this led to the introduction of the method called {\em multiscaling of low-variability periods} (MLVP) \cite{Kitt2005,Kalda2001}.

Let us consider a time-series $x(t)$, where $x$ can be, for example, a market index value. 
Then, a low-variability period of length $l_i$ is defined as a continuous time interval $T_i=[t_i,t_i+l_i]$ ($i=1,2,\ldots$) so that:
\begin{itemize}
 \item[(a)] 
\begin{equation} \label{delta}
|x(t) - \left<x(t) \right>_\tau| \le  \delta ~~~  \mbox{for} ~~~ t \in T_i
\end{equation}
and $\delta$ being a threshold parameter; angular braces denote sliding average over a window of width $\tau > \tau_0$ 
(in principle, the width of the window  can be arbitrarily large, however, the highest time resolution of the method 
and the widest scaling range is achieved when it is as small as possible, i.e., just a few cut-off scales $\tau_0$);
\item[(b)]
each period has maximal possible length, implying that decreasing $t_i$ or increasing $l_i$ would lead to violation of Eq.~(\ref{delta}). 
\end{itemize}
Note that in Eq.~(\ref{delta}), the left-hand-side can be alternatively normalized to the sliding average $\left<x(t)\right>_\tau$. 
This allows to improve the analysis of very long time-series, which are characterized by an exponential price growth. 
Further, we study the cumulative distribution function of the low-variability periods (the number of periods with $l_i \ge n$), $R(n)$.
We speak about multiscaling behaviour if there is a power law,
\begin{equation} \label{scal}
R(n)=R_0 \, n^{-\alpha(\delta,\tau)},
\end{equation}
where $\alpha(\delta,\tau)$ is a scaling exponent and $R_0$ is a constant.
Note that a particular case of this method (with $\tau$ being fixed to the cut-off scale $\tau_0$) has been introduced recently also in Refs.~\cite{Bogachev2007,Jung2008} and is referred to as the ``volatility return intervals analysis''. 

In Ref.~\cite{Kitt2009} the researchers of the Institute of Cybernetics showed theoretically that in the case of multifractal time-series, there is a MLVP-behaviour; this has been confirmed in Ref.~\cite{Bogachev2007}. 
Furthermore, it turns out that the scaling exponent $\alpha(\delta,\tau)$ is defined by the multifractal spectrum of the Hurst exponents $f(h)$ \cite{Kitt2005}: 
\begin{equation}\label{alpha}
\alpha(\delta, \tau) = f(\log_\tau  \delta).
\end{equation} 
Whereas the multifractal spectrum is a function of one variable, the MLVP exponent is a function of two variables, and hence, it can describe a wider class of scale-invariance. 
If the MLVP-behaviour is followed, Eq.~(\ref{alpha}) allows us to test the presence of multifractality. 
Indeed, if one plots $\alpha$ versus $\log_\tau  \delta$, the data for different values of $\tau$ and $\delta$ should lay on a curve, rather than being scattered over a plane. 
Using this method, it was found that currency exchange rates and market indices follow reasonably well 
multifractality for $\tau$ larger than a day, but fails at time scales smaller than a day \cite{Kitt2005}.
Let us note that for multifractal time-series the MLVP method effectively substitutes the multifractal scaling analysis, with two small benefits: it remains applicable when multifractal analysis fails and it tests the time-series at the highest available frequency [for $\alpha(\delta,\tau)$ one can use $\tau=\tau_0$, but for $f(h)$ any $h$ implies a scaling from the highest frequency down to lower frequencies, with the range of time scales representing at least a couple of octaves]. 

In Ref.~\cite{Kitt2005a} the method of MLVP analysis has been also applied to daily trading volumes for different stock indices. 
The multiscaling behaviour has been, indeed, observed, but there was no data collapse on the graph of scaling exponent $\alpha$ versus $\log_\tau  \delta$. 
Therefore, it was concluded that multifractality is not a good model for these data. 
Furthermore, the MLVP method  has been straightforwardly generalized to allow a multivariate data analysis. 
Upon introducing two thresholds, one for the volume and the other for the price movement, one can define the low variability periods as either the period during which none of the thresholds is exceeded, or the period during which at least one of the thresholds is exceeded. 
Not surprisingly, such a scaling analysis revealed a strong coupling of the price movements and the trading volumes \cite{Kitt2005a}.

Furthermore, it was shown that as soon as the price movements follow a MLVP behaviour, an interesting super-universality is to be expected: the probability of observing a larger-than-threshold movement (the ``silence-breaking'' probability) is inversely proportional to the length of the ongoing low-variability period. 
While theoretically this is a nearly obvious finding, it was not so easy to test on the basis of the financial data series: almost all these series are just too short to provide statistical volumes which are easy to analyse. 
In Ref.~\cite{Kitt2009} a novel data analysis technique was developed to overcome this problem of data sparseness. 
It allowed to confirm the validity of the super-universal scaling of the silence-breaking probability on the basis of various time-series.

A somewhat independent research subject that has been studied at the Institute of Cybernetics is directly related to the practical applications addressed in Ref.~\cite{kitt2004}: developing non-Gaussian portfolio optimization techniques. 
According to the efficient market hypothesis, the predictions of  the market price movements are bound to be well below the noise level. 
Any counterexample would be equivalent to violating the second law of thermodynamics; furthermore, it would cease to be valid as soon as you publish it. However, this fundamental restriction does not apply to the risk prediction. 
Therefore, it makes sense to try to improve the risk optimization techniques.

The problem of portfolio construction takes us back to the root of nonlinear time-series analysis and econophysics: what should the behaviour of investors in the financial markets be for combining the securities into the portfolio? 
The Nobel-prize-winning Markowitz has postulated (by assuming Gaussian movements in the financial time-series) that investors should use variance as their measure of  risk \cite{Markowitz1952}. 
The mean-variance optimisation is a centrepiece to a linear portfolio construction. 
However, the question can be raised: what is the true source of risk while investing into the stock markets? 
The central idea in Ref.~\cite{kitt2006} was that the risk can be divided into two parts: the Gaussian part and the leptokurtic part (the latter referring to the higher moments of the time-series). 
The correlation matrices of these two parts can differ significantly, and hence imply different optimal positions. 
So, the portfolio construction should first answer the question, which is the risk to be minimised. 
If it is a risk due to typical Gaussian price fluctuations then the mean-variance optimisation is a good approximation. 
However, if the main concern is to protect against the extreme (but rare) movements then another approach is to be taken. 
The simplest method is just to disregard the small Gaussian price fluctuations (below a threshold value, which serves as a control parameter) and devise the portfolio optimisation, based on the correlation matrix of the largest movements (corresponding to the ``fat tail'' of the distribution function). 
Of course, there is an additional problem: disregarding most of the data points reduces the reliability of the correlation matrix calculation. 
Therefore, it is important to avoid too high threshold parameter values and apply additional techniques (e.g. the factor analysis). 
Such a strategy has been implemented and analysed in Ref.~\cite{kitt2006} in the simplest case of two-asset portfolios. 
As expected, optimising against the non-Gaussian risk inflated slightly the standard deviation of the portfolios (as compared with the traditionally optimised portfolios), but the drawbacks (large and rapid losses of the portfolio value) were significantly smaller. 
Such studies cannot answer the most important question: should we reduce the small fluctuations, or the drawdowns? 
This is exclusively a subjective choice, but in some cases the answer is simple; it is provided by a (not perfect) parallel from the aviation industry: should we minimize the vibration in the cabin or the chances of falling down?


\section{Conclusion and outlook}


The present review of the econophysics research in Estonia is devoted to the 15th anniversary of the term ``econophysics''.
This context asks for a brief discussion of the history and translation of this word into Estonian language. 
The term {\em ``majandusf\"u\"usika''} was born in 2001, when R. Kitt started his PhD studies; thus,
it has  only a half of the age of the English word.
This noun is a straightforward compound of the words  ``majandus'', which means ``economics'', and ``f\"u\"usika'', which means ``physics''; 
therefore, the criticism which has been sometimes addressed to the English term is not applicable to the Estonian one.
However, in parallel also the term  ``\"okonof\"u\"usika'', adapted directly from English, has been used. 

In absolute numbers, the econophysics community in Estonia is small.
In fact, there are only 6 people who are working in this field, and thus far only one doctoral thesis \cite{KittPhD} has been defended in econophysics.
However, this makes 4 econophysicists per million habitants, which is no longer a small number.
Furthermore, we believe that many important results have been obtained.
The direct contacts between the scientists and the largest bank in Estonia open up a possibility to have an access to real data, 
otherwise often difficult to obtain.

Unfortunately, there is currently no tuition of econophysics at the Estonian universities; however, from time to time seminars are held.
Also, in year 2007 at the University of Tartu, two courses of complex systems were given by M.~Patriarca, where (among other topics) an overview of the basics concepts of econophysics was given.
Furthermore, one doctoral student is currently carrying out research under the supervision of R.~Kitt, 
in the Department of Economics at Tallinn University of Technology.
This allows us to believe that econophysics is a growing research field that surely has a future in Estonia.


\bibliography{./econphys15years}

\end{document}